\documentclass[10pt,letterpaper,bibnotes,notitlepage,final,balancelastpage,superscriptaddress,twocolumn,showpacs,pra]{revtex4}
\usepackage{amssymb}
\usepackage{amsmath}
\usepackage{amsfonts}
\usepackage{graphicx}
\usepackage{bm}
\usepackage{float}
\usepackage[bookmarks=false,pdfstartview=FitH,hyperindex=true, colorlinks, linkcolor=blue, citecolor=blue]{hyperref}
\usepackage[all]{hypcap}
\begin{document}
\title{Light pulse in $\Lambda$-type cold atomic gases}
\author{Ran Wei}
\affiliation{Hefei National Laboratory for Physical Sciences at Microscale and Department
of Modern Physics, University of Science and Technology of China, Hefei, Anhui
230026, China}
\author{Bo Zhao}
\email{bo.zhao@uibk.ac.at}
\affiliation{Institute for Theoretical physics, University of Innsbruck, A-6020 Innsbruck, Austria}
\affiliation{Institute for Quantum Optics and Quantum Information of the Austrian Academy
of Science, A-6020 Innsbruck, Austria}
\author{Youjin Deng}
\email{yjdeng@ustc.edu.cn}
\affiliation{Hefei National Laboratory for Physical Sciences at Microscale and Department
of Modern Physics, University of Science and Technology of China, Hefei, Anhui
230026, China}
\author{Shuai Chen}
\affiliation{Hefei National Laboratory for Physical Sciences at Microscale and Department
of Modern Physics, University of Science and Technology of China, Hefei, Anhui
230026, China}
\author{Zeng-Bing Chen}
\affiliation{Hefei National Laboratory for Physical Sciences at Microscale and Department
of Modern Physics, University of Science and Technology of China, Hefei, Anhui
230026, China}
\author{Jian-Wei Pan}
\affiliation{Hefei National Laboratory for Physical Sciences at Microscale and Department
of Modern Physics, University of Science and Technology of China, Hefei, Anhui
230026, China}

\pacs{32.80.Qk 42.50.Gy}

\begin{abstract}
We investigate the behavior of the light pulse in $\Lambda$-type cold atomic
gases with two counter-propagating control lights with equal strength
by directly simulating the dynamic equations and exploring the dispersion relation.
Our analysis shows that, depending on the length $L_0$ of the stored wave packet
and the decay rate $\gamma$ of ground-spin coherence, the recreated light can behave differently.
For long $L_0$ and/or large $\gamma$, a stationary light pulse is produced,
while two propagating light pulses appear for short $L_0$ and/or small $\gamma$.
In the $\gamma \rightarrow 0$ limit, the light always splits into two propagating pulses for
sufficiently long time.
This scenario agrees with a recent experiment [Y.-W. Lin, \emph{et al.,} Phys. Rev. Lett. \textbf{102},
213601(2009)] where two propagating light pulses are generated in laser-cooled
cold atomic ensembles.
\end{abstract}
\maketitle

\section{Introduction}

Quantum information transfer between light and atomic ensembles has attracted
much attention recently. In particular, electromagnetically induced
transparency (EIT) \cite{Harris,Fleischhauer2005}, a robust technique that renders
a resonant opaque medium transparent by means of destructive quantum
interference, has been exploited to realize the storage and retrieval of light
pulses in atomic ensembles \cite{Fleischhauer2000,Fleischhauer2002,Hau2001,Lukin2000}.
In the storage process, a weak probe light pulse carrying quantum
information and a strong coupling light are applied to
an optically thick atomic ensemble. The probe light is then
gradually converted into a ground-spin coherence as the coupling light
is adiabatically switched off, and,
as a result, the quantum information is stored in the atomic ensemble.
The reading process is almost reverse to the storage:
the control light is adiabatically turned on, and accordingly a new light pulse
is created and propagates out of the atomic ensemble. In this way,
the quantum information can be stored and read out without loss
in principle. If two counter-propagating control light pulses with equal
strength are adiabatically switched on~\cite{Andre2002}, the retrieved light pulse will
not propagate out of the atomic ensemble. Instead, it
stops in the media and forms a stationary light pulse, as
experimentally demonstrated
by Bajcsy \emph{et al.}~\cite{Bajcsy2003}.
On this basis, together with cold-atom techniques, many
applications are proposed, which
includes simulation of the dynamics of massive Schr\"odinger particles
~\cite{Fleischhauer2008}, of Dirac particles~\cite{Ottbach2009}, and
of the strong correlated Bose system confined in a hollow core fiber \emph{etc.}~\cite{Chang2008}.

The dynamic equations describing the behavior of the recreated light in $\Lambda$-type
atoms contain infinite-order terms. In conventional theoretical treatment,
secular approximation is used and only the zeroth-order coefficient of the ground-spin/optical coherence is kept~\cite{Zimmer2006}. In hot atomic systems like in Ref.~\cite{Bajcsy2003},
such a treatment is reasonable, since the higher-order terms decay very fast due to the random atomic motions and
collisions. However, thermal fluctuations are strongly
suppressed in cold atoms, and thus higher-order terms
decay much slower and should be considered. The light generation in cold
atomic ensembles has been theoretically studied beyond the secular
approximation by Hansen \emph{et al.}~\cite{Hansen2007}.
They came to the conclusion that, when the decay rate of the ground-spin/optical coherence is zero,
the generated light is a pure stationary light pulse-i.e., a stationary light without photon loss.
Nikoghosyan \emph{et al.}~\cite{Fleischhauer2009} took into account the relaxation of the upper state
and demonstrated that, under the slow-light condition,
the generated light is a stationary light pulse but with some photon loss.
Nevertheless, a recent experiment demonstrated~\cite{Liao2009} that, in laser-cooled cold atomic
ensembles, the retrieved light pulse is not stationary but
splits into two propagating wave packets.  The authors gave a simple model
which only involves the zeroth-order and first-order coefficients.
Since there is no obvious reason why the cutoff should take place at the first-order
term, our original minor motivation was to deal with the dynamic equations to
a higher order. It turns out that, in the zero-decay limit and under
adiabatic approximation, the dynamic equations can be analytically treated to
any order as one wishes (see Appendix for details).
Indeed, to the first order, one obtains
two counter-propagating light pulses, consistent with Ref. \cite{Liao2009}.
However, as higher-order terms gradually come in,
the relative group velocity of the two light pulses decreases and vanishes as
$\sqrt{2 \ell +1}/\ell$, where $\ell$ is the highest order of term in the calculation.
It seems to confirm the results in Refs. \cite{Hansen2007,Fleischhauer2009}.
Given this discrepancy and the potential important applications of the stationary light pulse
in cold atoms, a careful and systematic study seems desirable.

In this paper, we directly simulate the dynamic equations (given in Sec. II) to avoid further approximation.
For a given set of parameters, a series of simulations is performed with cutoff at
different order $\ell$ and the result are extrapolated to the infinite-order limit ($\ell \rightarrow \infty$).
These results are presented in Sec. III. Section IV provides a qualitative
understanding from the numerical calculation of the dispersion relation,
which is obtained from the Fourier transformation of the dynamic equations.
A brief discussion is given in Sec. V.

\section{Dynamic equations for the atom-light system}

\begin{figure}[ht]
\begin{center}
\includegraphics[width=8cm]{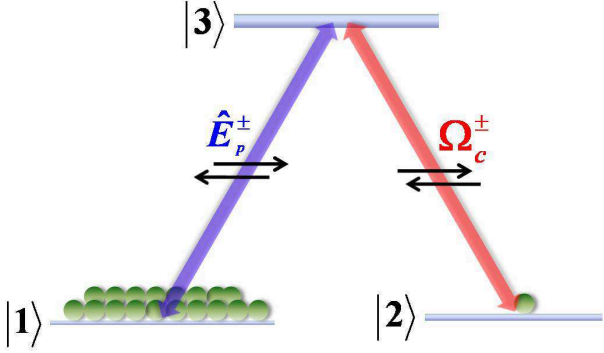}
\caption{
(Color online)
Sketch of the interaction between $\Lambda$-type atoms and
control light $\Omega_c^{\pm}$ and probe light $\hat{E}_p^{\pm}$.}%
\label{fig1}%
\end{center}
\end{figure}

Let us consider an ensemble of $\Lambda$-type atoms aligning along
a certain direction (say $z$), which is horizontal in Fig. \ref{fig1}. These
atoms interact with with a weak probe light $\hat{E}_p^{\pm}$
and a strong control light $\Omega_c^{\pm}$, treated as
quantum and classical light, respectively.

Under the single-mode approximation, the interaction Hamiltonian in the
rotating frame reads~\cite{Fleischhauer2002}

\[
{\cal H}=-\frac{N}{L}\int dz\hbar g\widetilde{\sigma}_{31}(z,t)\widetilde{E}%
_{p}(z,t)+\hbar\widetilde{\sigma}_{32}(z,t)\widetilde{\Omega}_{c}%
(z,t)+h.c.,
\]
where $g$ is the coupling constant, $\widetilde{\sigma}_{ij}(z,t):\equiv
\int dz_m\hat{\sigma}_{ij}^{m}\delta(z-z_m)$ is the continuous atomic operators,
with $\hat{\sigma}_{ij}^{m}:\equiv\left\vert i\right\rangle ^{m}\left\langle
j\right\vert $ the spin flip operator of the $m$th atom, $N$ is the atomic
number, $L$ is the length of the atomic ensemble, $\widetilde{E}%
_{p}(z,t)\mathbf{=}\hat{E}_{p}(z,t)e^{-i\omega_{p}t}$ is the electric field of
the probe light, and $\widetilde{\Omega}_{c}(z,t)\mathbf{=}\Omega
_{c}(z,t)e^{-i\omega_{c}t}$ is the Rabi frequency of the control field. For
simplicity, we assume the two ground states are degenerate, and the probe
light and control light are on resonance $\omega_{p}\approx\omega_{c}%
=\omega_{31}$. With slowly varying atomic operators
\begin{align*}
\hat{\sigma}_{13}(z,t)  &  =\widetilde{\sigma}_{13}(z,t)e^{i\omega_{p}t}\; ,\\
\hat{\sigma}_{12}(z,t)  &  =\widetilde{\sigma}_{12}(z,t)e^{i\omega
_{p}t-i\omega_{c}t} \; ,
\end{align*}
the Langevin equations governing the atomic dynamics read as
\cite{Scully}
\begin{align}
\frac{\partial\hat{\sigma}_{13}}{\partial t}  &  =-\Gamma\hat{\sigma}%
_{13}+ig\hat{E}_{p}+i\Omega_{c}\hat{\sigma}_{12}+\hat{F}_{13}\; ,%
\label{bloch1}\\
\frac{\partial\hat{\sigma}_{12}}{\partial t}  &  =i\Omega_{c}^{\ast}%
\hat{\sigma}_{13}+\hat{F}_{12}\; , \label{bloch2}%
\end{align}
where we have set $\hat{\sigma}_{11}=1$, $\hat{\sigma}_{33}
=\hat{\sigma}_{23}=0$. This approximation is appropriate since the probe field
is very weak and all the atoms are
initially prepared in $|1\rangle$. $\Gamma$ is the decay of the optical
coherence, and $\hat{F}_{ij}$ is the Langevin force.

The counter-propagating control light can be described by $\Omega
_{c}(z,t)=\Omega_{c}^{+}e^{ik_cz}+\Omega_{c}^{-}e^{-ik_cz}$, where
the light is assumed to be homogeneous. The probe light can also be
decomposed into two counter-propagating components as
\begin{equation}
\hat{E}_{p}(z,t)=E_{p}^{+}(z,t)e^{ik_cz}+E_{p}^{-}(z,t)e^{-ik_cz} \; .\label{probe}%
\end{equation}
Following the standard procedure~\cite{Zimmer2006,Hansen2007,Fleischhauer2009}, we define
ground-spin coherence as $S=\sqrt{N}\hat{\sigma}_{21}$ and optical coherence as
$P=\sqrt{N}\hat{\sigma}_{31}$, and expand them as
\begin{align}
S &  =\sum_{n=-\infty}^{n=\infty}S_{2n}e^{2nik_cz}\label{expansion1} \; ,\\
P &  =\sum_{n=-\infty}^{n=\infty}P_{2n+1}e^{(2n+1)ik_cz} \; .\label{expansion2}%
\end{align}
Inserting Eqs.(\ref{probe})-(\ref{expansion2}) into Eqs.(\ref{bloch1}) and
(\ref{bloch2}) and assuming $\Omega_c^{\pm}=\Omega_c$,
we obtain a set of dynamic equations as
\begin{align}
\frac{\partial P_{2n+1}}{\partial t} &  =-(\Gamma+\gamma_{2n+1})P_{2n+1}%
\nonumber \; ,\\
&  +ig\sqrt{N}E_{p,2n+1}+i\Omega_{c}(S_{2n}+S_{2(n+1)})\label{infinite1}\\
\frac{\partial S_{2n}}{\partial t} &  =-\gamma_{2n}S_{2n}+i\Omega_{c}%
(P_{2n-1}+P_{2n+1}) \; ,\label{infinite2}%
\end{align}
where $E_{p,\pm1}=E_{p}^{\pm}$, $E_{p,2n+1}(n\neq0,-1)=0$, and
$\gamma_n$ represents the decay of $n$th-order coefficient.
We have also neglected the Langevin force terms since they do not play a role
in the long-time behavior of the light pulse.
The dynamics of the probe light is governed by the Maxwell equations
\begin{align}
\frac{\partial E_{p}^{+}}{\partial t}+c\frac{\partial E_{p}^{+}}{\partial z}
&  =ig\sqrt{N}P_{1}\label{max1} \; ,\\
\frac{\partial E_{p}^{-}}{\partial t}-c\frac{\partial E_{p}^{-}}{\partial z}
&  =ig\sqrt{N}P_{-1} \; .\label{max2}%
\end{align}

For warm atomic vapors, the random motions and collisions of atoms
result in a very rapid decay of spatial coherence. Effectively,
one has $\gamma_0 =0$ and $\gamma_n >0$ for $n \neq 0$.
In this case, multiple components $S_{2n}$($n\neq 0$) of ground-spin coherence are suppressed,
the same applies to $P_{2n+1}$($n \neq -1, 0$).
Thus, these terms can be neglected, and the probe light forms
a stationary light pulse~\cite{Zimmer2006}.

\begin{figure}[ht]
\begin{center}
\includegraphics[width=9cm]{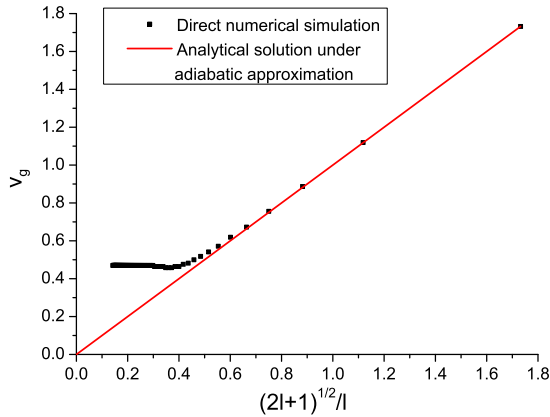}
\caption{
(Color online)
Group velocity $v_g$ of the recreated forward (backward)-propagating light pulse
in the $\gamma_n=0$ limit. The unit of $v_g$ is $\frac{c\Omega_c^2}{g^2N}$.
The red straight line is from the approximate analytic
calculation (\ref{wave1})(\ref{wave2}) in Appendix, while the
black dots are obtained by the direct numerical simulation
of Eqs.(\ref{infinite1})-(\ref{max2}) in Sec. III. The error margins of the
data points are $\pm 0.006$.}
\label{fig2}%
\end{center}
\end{figure}

\begin{figure}[ht]
\begin{center}
\includegraphics[
width=8cm]{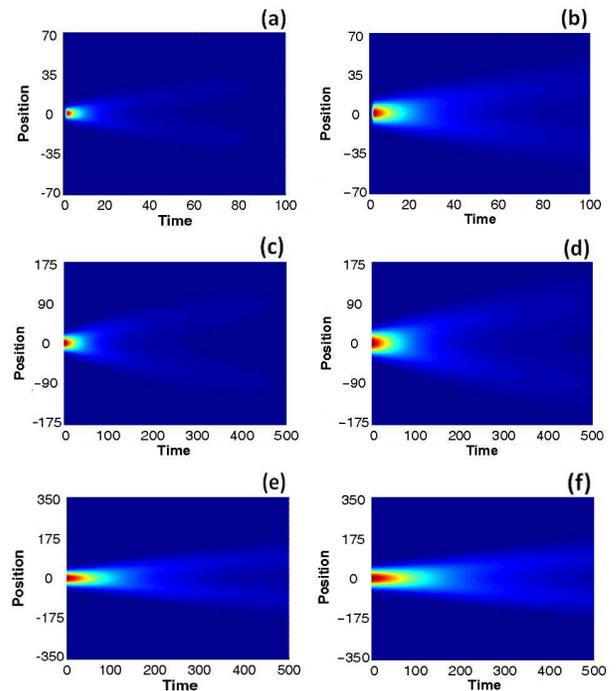}
\caption{
(Color online)
Light intensity $|E_{s}|^{2}+|E_{d}|^{2}$ as a function of time and position $z$
for $\gamma_n=0$ with different length $L_{0}$. Figures (a)-(f) represents $L_0/l_{abs}=5, 10, 20, 30, 40$, and 50,
respectively. The position is in unit of
$l_{abs}$ and the time is in unit of $1/\Gamma$. Strong light is shown in red bright color,
while the back ground is in blue.}%
\label{fig3}%
\end{center}
\end{figure}

In contrast, in a deep optical lattice, where
atoms are fixed at the lattice sites~\cite{Bloch}, the decays of
the higher-order coefficients can be ignored.
In other words, one has $\gamma_{n} = 0$ for any $n$. Thus,
the multiple components can be populated and preserve their
coherence~\cite{Zimmer2006}, and secular approximation is no longer valid.
Nevertheless, after adiabatic elimination, one can
analytically solve the Eqs.(\ref{infinite1})--(\ref{max2}) with $\gamma_n=0$
(see Appendix for details). The group velocity $v_g$
of the forward (backward)-propagating light pulse is shown in Fig.\ref{fig2}.
One finds that: 1), the approximate solution with cutoff at finite $\ell$
yields a non-zero $v_g$, and 2), $v_g (\ell)$ reaches the maximum value
at $\ell =1$, and then vanishes with $\sqrt{2 \ell+1}/\ell$.
Since the real system corresponds to the infinite-order limit, one ``seems"
to conclude that the recreated light forms a stationary pulse.

For cold atomic systems where temperature is low but nonzero,
one has $\gamma_0=0$ and $\gamma_n \neq 0$
for $n \neq 0$.  In this case, we set $\gamma_{n}=|n|a\Gamma$, with $a$
the decay constant. This decay model can well describe the decays of the coefficients of the
ground-spin/optical coherence in various cold atomic systems.
For instance, in the laser-cooled cold atomic ensembles,
the higher-order coefficients have a phase grating of $e^{ink_cz}$
across the atomic gases, and thus will decay due to atomic random motion. The
decay rate can be estimated by the time needed for the atoms moving across one
wavelength of the phase grating $\gamma_{n}\sim|\frac{v_{s}}{\lambda}%
|=n|\frac{k_cv_{s}}{2\pi}|$ \cite{Zhaobo}. In the Bose condensation, the
higher-order coefficients can be regarded as a particle excitation with momentum $|n|k_c\hbar$,
At average, they move out of the atomic gases after a time of $L(\frac{|n|k_c\hbar}%
{m})^{-1}$ \cite{Hau2007}, and the decay rate can be approximated by
$\gamma_{n}$ $\sim\frac{|n|k_c\hbar}{mL}$.

\begin{figure} 
\begin{center}
\includegraphics[width=8cm]{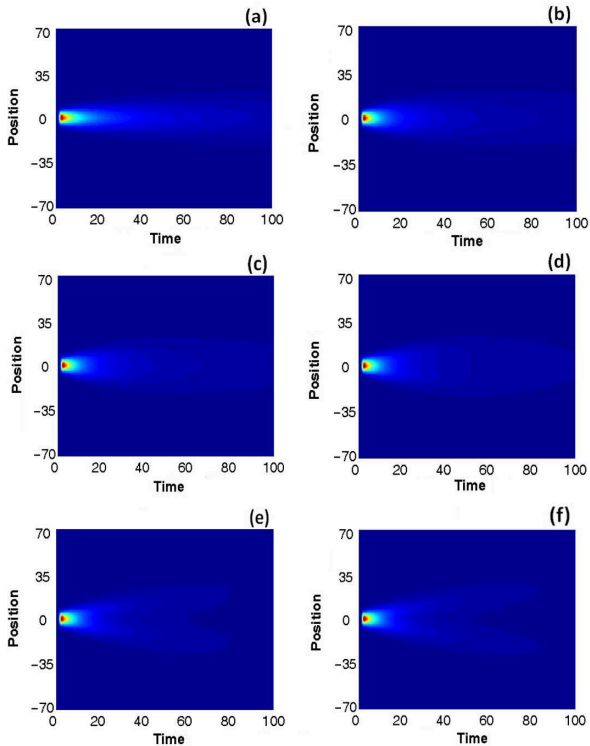}
\caption{
(Color online)
Light intensity $|E_{s}|^{2}+$ $|E_{d}|^{2}$ for $L_{0}%
=5 l_{abs}$ with different decay constant $a$.  Figures (a)-(f) are for $a=0.2,
0.02, 0.01, 0.005, 0.001$, and $0$, respectively. The position is in unit of
$l_{abs}$ and the time is in unit of $1/\Gamma$.}%
\label{fig4}%
\end{center}
\end{figure}

\begin{figure}
\begin{center}
\includegraphics[width=8cm]{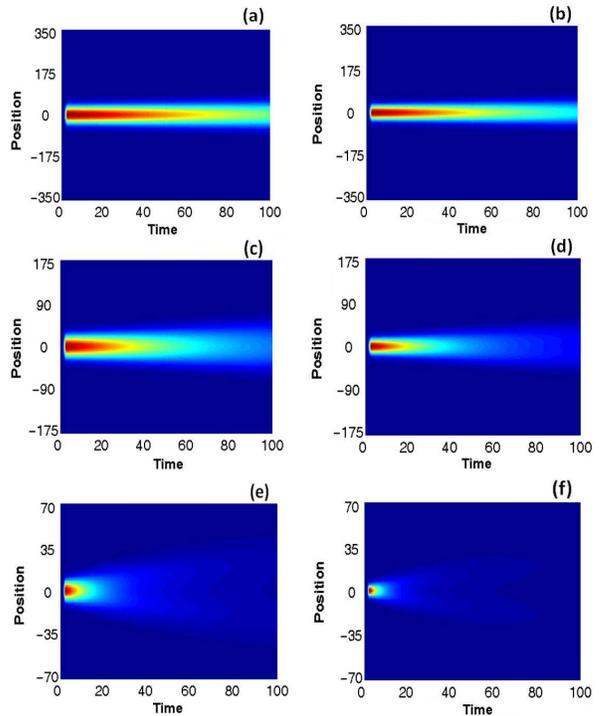}
\end{center}
\caption{
(Color online)
Light intensity $|E_{s}|^{2}+$ $|E_{d}|^{2}$ for $a=0.001$ with
different length $L_{0}$. Figures (a)-(f) are for $L_0/l_{abs}=50, 40, 30, 20, 10$, and $5$,
respectively. The position is in unit of $l_{abs}$ and the time is in unit of $1/\Gamma$.}%
\label{fig5}%
\end{figure}

\section{Numerical solution}

To check the validity of the approximate solution in Fig.~\ref{fig2}
and further find the dynamics of the recreated light pulse in cold
atomic systems ($\gamma_n =|n| a \Gamma$) for which the approximate treatment is
unavailable, we directly simulate
Eqs.(\ref{infinite1})--(\ref{max2}). Naturally, a cutoff takes place
at finite $\ell$ and accordingly $ 5 + 4 \ell$ equations are
involved in each simulation. The result for real systems ($ \ell \rightarrow \infty$)
is obtained from the extrapolation of simulations for finite $\ell$.

We first consider the zero-decay limit ($\gamma_n=0$). The initial condition is taken such that:
1), only the zeroth component $S_0 (z,t=0)$ of the ground-spin coherence
is nonzero while all other components $S_{2n} (z,0)$ are zero; $S_0$ assumes a Gaussian
shape $S_{0} (z,0)=e^{-(z/L_{0})^{2}}$ with $L_0$ the length of the wave packet;
2), all components
of the optical coherence are zero $P_{2n+1}(z,0)=0$, and 3), no probe light
exists at the beginning $E_{p}^{+}(z,0)=E_{p}^{-}(z,0)=0$.
Further, the wave-packet length is set at $L_0=5 l_{abs}$,
with $l_{abs}=\frac{\Gamma c}{g^{2}N}$ the absorption length.
To be in the slow-light regime, we chose the parameters to be
$\Omega_c=0.69\Gamma$, $g^2N=138\Gamma^2$.
The equations are directly solved by Lax-Friedrichs method with sufficiently small step.
The simulation is up to $\ell =100$, and the group velocity $v_g$
of the recreated forward (backward)-propagating light pulse is measured. The results are shown in Fig.~\ref{fig2}.
As the approximate analysis, $v_g (\ell)$ reaches its maximum at $\ell=1$
and then starts to decrease. Nevertheless, the decrease of $v_g$ becomes slower and
slower after $\ell \approx 14$ and eventually stays unchanged at
$v_g = 0.47 \pm 0.006$. From Fig.~\ref{fig2}, it looks rather secure to conclude
that the group velocity $v_g$ takes a finite value for $\ell \rightarrow \infty$,
suggesting that the recreated light splits into two counter-propagating light
pulses. This defies the earlier approximate solution that the light forms a
stationary light pulse.

We then consider the effect of the wave-packet length $L_0$ for the $\gamma_n =0$ case
and set $L_0/l_{abs} =5, 10, 20, 30, 40$, and $50$.
We define the sum mode $E_{s}=E_{p}^{+}+E_{p}^{-}$ and difference mode $E_{d}=E_{p}%
^{+}-E_{p}^{-}$, and measure the intensity of the light pulses $|E_{s}|^{2}+$ $|E_{d}|^{2}$
as a function of time and position $z$. The results for $\ell =30$,
where the simulation already reaches the steady state, are shown in
Fig.~\ref{fig3}.
One observes that, for all the cases, the generated light always splits into two
counter-propagating light pulses. Nevertheless, as $L_0$ grows, the time for
the occurrence of splitting becomes longer and longer.
In experiments that are finished within short times, one may not be able to
observe such a splitting.

Next, we study the decay model for cold atomic systems ($\gamma_n =|n| a \Gamma$).
Shown in Figs. \ref{fig4} (a)-(f) are the dynamics for $L_0=5l_{abs}$ with
$a =0.2, 0.02, 0.01, 0.005, 0.001$ and 0.
When the decay rate is large--i.e., $a$ is large,
the generated light forms a stationary light pulse with dissipating; see
Fig.\ref{fig4}(a)-(c). In contrast, for sufficient small $a$ (Fig.\ref{fig4}(d)-(f)),
two counter-propagating light pulses appear.

Figures \ref{fig5} (a)-(f) display the dynamics of the generated light
for $a=0.001$ with $L_0/l_{abs}=50, 40, 30, 20, 10$, and $5$.
It can be seen that the generated light is stationary for long wave length $L_0$
while splits into two light pulses for small $L_0$. We should mention
that, however, the dynamics shown in Fig.\ref{fig5} is up to time
$100/\Gamma$, much shorter than $500/\Gamma$ in Fig.\ref{fig3}(c)-(f).
To see the behavior of the light pulse for longer time, we
performed the calculation for $L_0/l_{abs}=50$ up to the time $500/\Gamma$ and
did not observe any splitting.

\begin{figure}
\begin{center}
\includegraphics[width=10cm]{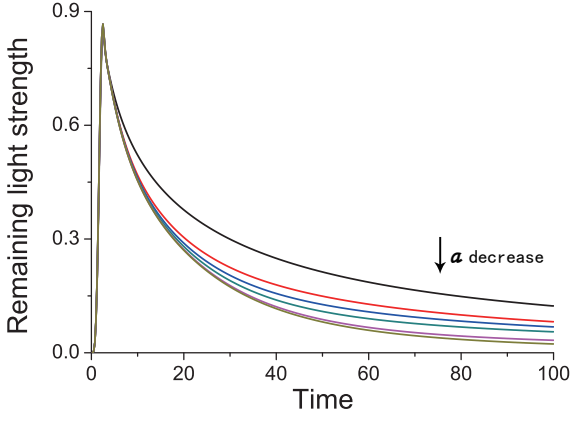}
\caption{
(Color online)
Remaining light strength $I$ as a function of time for $L_{0}=5l_{abs}$ with
$a=0.2$, $a=0.02$, $a=0.01$, $a=0.005$, $a=0.001$, and $a=0$.
The unit of $I$ is arbitrary and the unit of time is $1/\Gamma$.}%
\label{fig6}%
\end{center}
\end{figure}

We also calculated the total light strength $I=\int(|E_{s}|^{2}+|E_{d}|^{2})dz$
remaining in region $\{-3L_{0},3L_{0}\}$ as a function of time.
The results for $L_0=5l_{abs}$ and different $a$ are shown in Fig.\ref{fig6}.
Indeed, as $a$ decreases, the remaining light strength $I$ becomes weaker and weaker,
reflecting that the loss of photons becomes more and more serious due to the propagating
of the splitting light pulses.

\section{Dispersion relation}

The earlier simulation yields directly observable phenomena. In this section, we
aim to provide a qualitative understanding by exploring the associated
dispersion relation.

Note that Eqs.(\ref{infinite1})-(\ref{max2}) are linear,
the Fourier transformation of these equations leads to

\begin{align}
(\Gamma+\gamma_{|2n+1|}-i\omega)P_{2n+1}  &  =ig\sqrt{N}E_{p,2n+1}\nonumber\\
&  +i\Omega_{c}(S_{2n}+S_{2(n+1)}) \; , \label{infinite5}\\
(\gamma_{|2n|}-i\omega)S_{2n}  &  =i\Omega_{c}(P_{2n-1}+P_{2n+1}%
) \; , \label{infinite6}\\
-i\omega E_{p}^{+}+ickE_{p}^{+}  &  =ig\sqrt{N}P_{1} \; , \label{infinite7}\\
-i\omega E_{p}^{-}-ickE_{p}^{-}  &  =ig\sqrt{N}P_{-1}  \; . \label{infinite8}%
\end{align}

After some tedious calculations, similar as those in Appendix,
we obtain the dispersion relation%
\begin{align}
k =\pm\frac{i\Gamma}{c}\sqrt{(\frac{\frac{g^{2}N}{\Gamma^{2}}}{\frac
{\mathit{\Gamma}_{s}(\omega)}{\Gamma}}-\frac{i\omega}{\Gamma})(\frac
{\frac{g^{2}N}{\Gamma^{2}}}{\frac{\mathit{\Gamma}_{d}(\omega)}{\Gamma}}%
-\frac{i\omega}{\Gamma})},
\label{dispersionexpression}
\end{align}
where we have introduced two effective decay parameters $\mathit{\Gamma}_{S}(\omega)$
and $\mathit{\Gamma}_{D}(\omega)$

\begin{align}
\mathit{\Gamma}_{s}(\omega)  &  =\Gamma+\gamma_{1}-i\omega+\frac{2i\Omega
_{c}^{2}}{\omega}+{\mathcal R} \; ,
\nonumber\\
\mathit{\Gamma}_{d}(\omega)  &  =\Gamma+\gamma_{1}-i\omega+ {\mathcal R} \; ,
\nonumber
\end{align}
with parameter ${\mathcal R}$ recursively expressed as
\[
{\mathcal R} :\equiv \frac{\Omega_{c}^{2}}{\gamma_{2}-i\omega+\frac{\Omega_{c}^{2}}
 {...+\frac{\Omega_{c}^{2}}{\Gamma+\gamma_{2\ell-1}-i\omega+\frac
 {\Omega_{c}^{2}}{\gamma_{2\ell}-i\omega...}}}} \; .
\]
In the dispersion relation~(\ref{dispersionexpression}), the real and the imaginary parts
of momentum $k$, $|$Re$(k)|$ and $|$Im$(k)|$, qualitatively characterize
the effects of dispersion and of dissipation, respectively.

\begin{figure}[ht]
\includegraphics[width=8.0cm]{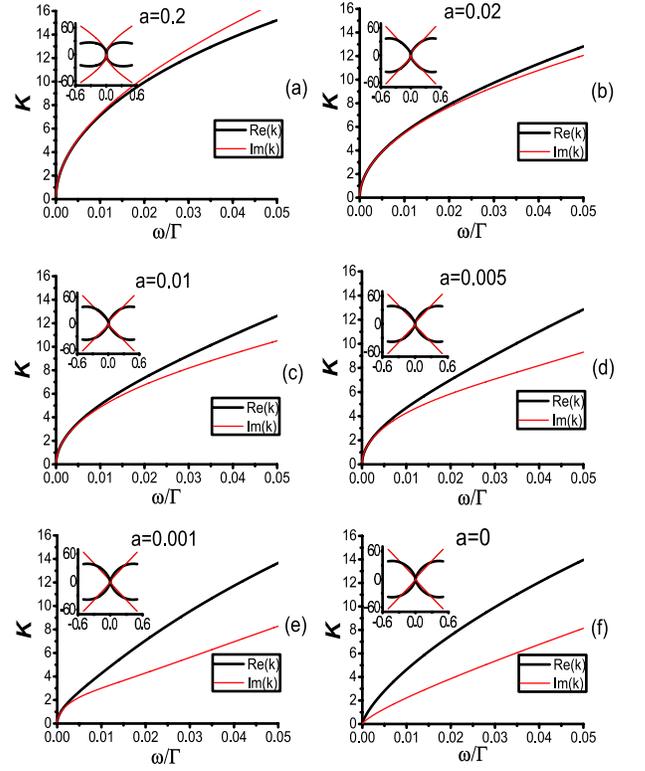}
\caption{
(Color online)
Numerical results for dispersion
relation~(\ref{dispersionexpression}) for $L_0 =5 l_{abs}$ and different decay constants $a$.
The vertical axis represents the real and the imaginary part of momentum
$k$: $K:\equiv (c/{\Gamma})$Im($k$) (the red thin curve)
and $(c/{\Gamma})$Re($k$) (the black thick curve). The insets display the dispersion relation in
a much larger region.}%
\label{fig7}%
\end{figure}

For a given frequency $\omega$, we numerically calculate Re$(k)$ and Im$(k)$
by taking a cutoff at $\ell$ for ${\mathcal R}$, and then
extrapolate the calculations to $\ell \rightarrow \infty$.
The parameters $g^{2}N=138\Gamma^{2}$ and $\Omega_c=0.69\Gamma$ are the
same as in Sec. II, and the decay model $\gamma_{n}=|n|a\Gamma$ is used.
Figure~\ref{fig7} displays the results for $L_0 =5 l_{abs}$ at $\ell=1000$,
where ${\mathcal R}$ already reaches its steady value.
In Fig.\ref{fig7}(a) where $a=0.2$ is comparable to $1$, the black thick (Re$(k)$) and
the red thin (Im$(k)$) curves almost overlap in the vicinity of original point $(|k|=0,\omega=0)$.
As $\omega$ increases, they gradually separate from each other.
The red thin curve increases faster than the black thick one--i.e., $|$Im$(k(\omega))$$|$$>$$|$Re$(k(\omega))$$|$.
This implies that dissipation dominates over dispersion.
When $a$ becomes smaller and smaller, the two curves separate at smaller frequency $\omega$.
Furthermore, one has $|$Im$(k(\omega))$$|$$<$$|$Re$(k(\omega))$$|$.
Namely, the effect of dispersion becomes more important than that of dissipation.
For $a<0.01$ (Fig.\ref{fig7}(d)-\ref{fig7}(f)), the two curves
significantly separate from each other as long as $|k|$ deviates
from zero. This means that the light pulse has a nonzero group velocity
and can propagate out of the atomic ensembles. In addition, one observes that the black curve grows less and
less rapidly as $a$ decreases, reflecting that the effect of dissipation becomes
weaker and weaker. The insets show that, for very large $|k|$, the dissipation always play the major role.

In short, we argue that the phenomena in Sec. III can be qualitatively understood
from the competition between the effects of dispersion and of dissipation.
For $a \approx 1$, the dissipation dominates over the dispersion,
and a stationary light pulse is generated. This applies to thermal atomic gases.
For $a \ll 1$, the dispersion wins as long as $|k| > 0$.
If the length of the stored wave packet $L_0$ is sufficiently long--i.e., $|k| \approx 0$,
the recreated light forms a stationary pulse; otherwise, it splits into
two counter-propagating light pulses.
In the $a=0$ limit and for a finite length wave packet, the generated light will always splits
into two propagating light pulses for sufficiently long time, since the dispersion is always dominant
over dissipation.

\section{Discussion}

In summary, using direct simulation of the dynamic equations for $\Lambda$-type atomic
systems, we find that both the decay rate $\gamma$ and the length $L_0$ of the
stored wave packet play an important role in determining the behavior of the new
light generated by two counter-propagating control lights with equal strength.
The numerical simulation of the $\gamma=0$ limit defies the approximate analytical solution.
This means that the adiabatic-elimination treatment demonstrated in Appendix is invalid.
For cold atomic systems, our calculations suggest that the recreated light
forms a stationary pulse for large $L_0$ and/or $\gamma$ while splits into
two counter-propagating light pulses for small $L_0$ and/or $\gamma$.
This scenario agrees well with the recent experiment. A qualitative understanding is
given from the aspect of the dispersion relation. We expect that our systematic calculation
shall provide useful information for future experiments.

\acknowledgments
We are very grateful to Ite A.Yu and Tao Xiong for helpful discussions.
This work is supported by the NNSFC, the NNSFC of Anhui (under Grant No. 090416224),
the CAS, and the National Fundamental Research Program
(under Grant No. 2006CB921900).

\section*{Appendix}

For $\gamma_n=0$, Eqs.(\ref{infinite1}) and (\ref{infinite2}) can analytically
solved under additional approximation--i.e., the adiabatic elimination,
as illustrated below.
Assuming that the characteristic interaction time $T$ is long compared to
the upper level relaxation--i.e., $\frac{1}{\Gamma}\ll T$,
we can adiabatically eliminate $\frac{\partial P_{2n+1}}{\partial t}$ in
Eqs.~(\ref{infinite1}) and (\ref{infinite2}) and obtain
\begin{eqnarray}
   \label{infinite1'}
        \Gamma P_{2n+1}= igE_{p,2n+1}+i\Omega_c(S_{2n}+S_{2(n+1)}) \; ,\\
   \label{infinite2'}
        \frac{\partial S_{2n}}{\partial t} = i\Omega_c(P_{2n-1}+P_{2n+1}) \; .
\end{eqnarray}

\

Differentiating both sides of Eq.~(\ref{infinite1'}) with respect to $t$ and
substituting Eq.~(\ref{infinite2'}) yield
\begin{eqnarray}
   \label{infinite3'}
        \notag \Gamma\frac{\partial P_{2n+1}}{\partial t}= ig\frac{\partial E_{p,2n+1}}{\partial t}\\
        -\Omega_c^2(P_{2n-1}+2P_{2n+1}+P_{2n+3}) \; .
\end{eqnarray}

Once more, the derivative terms can be eliminated by adiabatic treatment, and
the equations become
\begin{eqnarray}
   \label{s2}
   P_{s,2n+1}=(-1)^{n}((2n+1)P_{s,1}-n\frac{i g}{\Omega_c^2}\frac{\partial E_s}{\partial t}) \; ,\\
   \label{d2}
   P_{d,2n+1}=(-1)^{n}(P_{d,1}-n\frac{i g}{\Omega_c^2}\frac{\partial E_d}{\partial t}) \; .
\end{eqnarray}
where we have introduced s mode and d mode
\begin{eqnarray}
P_{s,2n+1}&=&P_{2n+1}+P_{-(2n+1)} \; ,\\
P_{d,2n+1}&=&P_{2n+1}-P_{-(2n+1)} \; .
\end{eqnarray}

Neglecting terms $P_{s,2n+1}$ and $P_{d,2n+1}$ for $n \ge \ell$,  one has
\begin{eqnarray}
   \label{final1}
   P_{s,1}=\frac{\ell}{2\ell+1}\frac{i g}{\Omega_c^2}\frac{\partial E_s}{\partial t} \; ,\\
   \label{final2}
   P_{d,1}=\ell\frac{i g}{\Omega_c^2}\frac{\partial E_d}{\partial t} \; .
\end{eqnarray}

Making use of the Maxwell equations and the initial condition $S_n =0$ for $n \neq 0$,
we obtain the propagating solution
\begin{small}
\begin{eqnarray}
   \label{wave1}
   E_s(z,t)=-(\frac{\Omega_cS_{0}(z-c_0t,0)}{g}
   +\frac{\Omega_cS_{0}(z+c_0t,0)}{g}) \; ,\\
   \label{wave2}
   E_d(z,t)=-\frac{\rho_0}{c_0}(\frac{\Omega_cS_{0}(z-c_0t,0)}{g}
   -\frac{\Omega_cS_{0}(z+c_0t,0)}{g}) \; .
\end{eqnarray}
\end{small}

Where the group velocity of the splitting wave packet is
$c_0=c/\sqrt{(1+\frac{\ell g^2N}{(2\ell+1)\Omega_c^2})(1+\frac{\ell g^2N}{\Omega_c^2})}$.
In low the group-velocity limit, $c_0\approx c\frac{\sqrt{2\ell+1}\Omega_c^2}{\ell g^2N}$.\\

\end{document}